\documentclass{article}

\DeclareMathAlphabet{\altmathcal}{OMS}{cmsy}{m}{n}
\usepackage[english]{babel}

\usepackage{tikz}
\usepackage{caption}
\usepackage[letterpaper,top=2cm,bottom=2cm,left=3cm,right=3cm,marginparwidth=1.75cm]{geometry}
\usepackage{cite}
\usepackage{amsmath}
\usepackage{amssymb}
\numberwithin{equation}{section}
\usepackage{graphicx}
\usepackage[colorlinks=true, allcolors=blue]{hyperref}
\usepackage{appendix}

\usepackage{authblk} 

\title{Deriving Tsallis entropy from $q$-deformed Hamiltonian within a statistical mechanics framework}
\author{Paradon Krisut$^{1,\dagger}$ and  Sikarin Yoo-Kong$^{2,\ddagger}$ \\
\small {$^1$Department of Physics,}
\small \emph{ Mahidol University, Bangkok, Thailand, 10400.}\\ 
            \small {$^2$The Institute for Fundamental Study (IF),} \small\emph{Naresuan University, Phitsanulok, Thailand, 65000.}\\
		\small{$^\dagger$paradon.kri@student.mahidol.edu, $^\ddagger$sikariny@nu.ac.th} \\
	}
\usepackage{graphicx}

\begin{document}
\date{}
\maketitle
\begin{abstract}
%
The Tsallis entropy, which possesses \(q\)-additive property, is derived from the first principle employing the non-additive Hamiltonian or the $q$-deformed Hamiltonian with the canonical ensemble assumption in statistical mechanics. Here, the $q$-algebra and properties of $q$-deformed functions are extensively used throughout the derivation.  Consequently, the thermodynamic quantities, e.g. internal energy and Helmholtz free energy, are derived and they inherently exhibit the \(q\)-additivity. From this intriguing connection between Tsallis entropy and the $q$-deformed Hamiltonian, the parameter $q$ encapsulates the intrinsic degree of \(q\)-additivity for the thermodynamic systems.
\end{abstract}

\section{Introduction}
\label{intro}
In 1988, Tsallis introduced a new form of entropy, later known as the Tsallis entropy, as a generalization of Boltzmann-Gibbs entropy \cite{Tsallis1988}. A key feature of Tsallis entropy is its \textcolor{black}{non-additive (pseudo-additive) property, which we shall refer to as $q$-additivity (see Section \ref{q-H})}, governed by the parameter $q$, which makes it suitable for describing complex systems beyond the reach of standard entropy. Since then, Tsallis entropy has been extensively researched and continuously developed. Through the maximal entropy principle (MSE) concept \cite{PhysRev.106.620}, Tsallis entropy has been extensively investigated and applied in a wide range of fields, including engineering, physics, other sciences, information theory, and economics.
\textcolor{black}{In physics, Tsallis entropy has particularly made a huge impact in statistical mechanics, plasma physics \cite{TARUYA2003285,Jiulin2007,e13101765,plastino1999,DU2004262,PhysRevE.78.040102,Livadiotis_2010,PAVLOS2015113,e21090820, DEALMEIDA2005395, MARTINEZ1998183, e14040701, e13111928 }, quantum information \cite{PhysRevA.63.042104, ABE2001157, PhysRevLett.88.170401, CANOSA2005121, KHORDAD2017559, Zhao2018, PhysRevE.101.040102} and black hole thermodynamics \cite{Tsallis2013,e22010017,LUCIANO2023101319}. }
Beyond physics\footnote{The comprehensive overview is given in \cite{Tsallis2023ch7}.}, Tsallis entropy has been applied to various complex systems, including those with non-Markovian dynamics or fractal structures \cite{PhysRevE.53.4754, cáceres_1999, TSALLIS2002371, CASTRO2005184}, and financial markets \cite{e22040452}. However, Tsallis entropy remains a subject of ongoing debate, particularly regarding the interpretation and physical significance of the entropic index \(q\). Some researchers have attempted to provide context-specific interpretations, suggesting that $q$ may represent the degree of correlation or interaction strength within a system \cite{DEALMEIDA2005395,e13101765, Tsallis2009, HANEL2005260, Kim2019, PhysRevE.101.040102, Wang2011, OU20085761, Jiang2012}. But, these interpretations are typically tailored to particular physical contexts and thus lack universal applicability. Therefore, this limitation underscores the persistent debate over the fundamental significance and versatility of Tsallis entropy in physics. 
\\
\\
In recent years, advances in our understanding of Lagrangian non-uniqueness have led to the concept of multiplicative Lagrangians for systems with a single degree of freedom, giving rise to what is termed as the one-parameter Lagrangian \cite{MultiL}. Intriguingly, this new type of Lagrangian gives an infinite hierarchy of Lagrangians, producing the same equation of motion. This approach has proven to be effective in field theory applications, providing promising solutions to long-standing challenges such as the Higgs hierarchy problem \cite{Supanyo2022} and the neutrino mass problem \cite{Supanyo:2023jkh}. In the Hamiltonian context, one can apply the Legendre transformation leading to the multiplicative Hamiltonian (or one-parameter Hamiltonian). Again, this new Hamiltonian gives an infinite hierarchy of Hamiltonians, producing the same equation of motion. A key interesting feature of this new Hamiltonian is the \(q\)-additive property—reminiscent of that of the Tsallis entropy—suggesting a possible connection between the two frameworks. So far, in the literature, one finds that there were attempts to construct the statistical ensemble through the Tsallis entropy with the standard Hamiltonian leading to some restrictions on defining the temperature \cite{DEALMEIDA2005395, HANEL2005260} and including the interaction into the system \cite{OU20085761, Jiang2012}\textcolor{black}{, see also  \cite{PLASTINO1994140, PARVAN2006331,Marino2007}
}. Then, in this work, we shall use the $q$-deformed Hamiltonian to derive the Tsallis entropy within the statistical mechanics framework without relying on any prior assumptions.
\\
\\
The paper is organised in the following. In section \ref{q-H}, the $q$-deformed Hamiltonian will be briefly discussed together with similar properties to the Tsallis entropy. In section \ref{Construct_ensemble}, the phase space density matrix, for the microcanonical and canonical ensembles, associated with the $q$-deformed Hamiltonian is systematically constructed. The thermal equilibrium condition is also discussed. In section \ref{Entropy_Thermal_equilibrium}, the candidate entropy function is derived based on the first and second laws of thermodynamics, ultimately yielding a form equivalent to Tsallis entropy. In section \ref{Connection_w/_non_extensive_thermo}, the non-additive thermodynamics will be examined. The $q$-version of the internal energy and Helmholtz's free energy are given as well as their \(q\)-additive property. In section \ref{second_entropy}, on requiring the consistent definition of internal energy from the previous section, the candidate entropy function will be reformulated using a revised notion of the phase space density matrix leading to a perfect match with the Tsallis entropy. In section \ref{summary}, a summary is provided together with a discussion on reinterpreting the parameter $q$ as a measure of \(q\)-additivity rather than correlation.

\section{The $q$-deformed Hamiltonian}
\label{q-H}
In this section, we will give a brief review of the multiplicative Lagrangian and Hamiltonian and also will point out how to construct the $q$-deformed Hamiltonian from the multiplicative one as well as an intriguing similar structure between the $q$-Hamiltonian and the Tsallis entropy.
\\
\\
To make this section consistently clear, let us first review some features of the Tsallis entropy given by
\begin{equation}\label{Tq}
T_q \, = \, -k \, \frac{1-\sum_i p_i^q}{1-q} \, , \hspace{0.5cm} q \in \mathbb{R},
\end{equation}
where \( q\) is the parameter and \(k\) is the Boltzmann constant. The key feature of Tsallis entropy is the non-additive property, also called the pseudo-additivity. If system 1 and system 2 are independent, the joint probability satisfies the product rule \(\, p_{ij}^{1 2} \, = \, p_i^1 p_j^2 \, \) resulting in
\begin{equation}
    T_q^{12} = T_q^1 + T_q^2 + (1-q)\frac{T_q^1 T_q^2}{k}
    \label{pseudo_additive_of_T_q} \;.
\end{equation}
 \textcolor{black}{Here, \( \, T_q^{12}, T_q^1\) and \(T_q^2 \, \) represent the Tsallis entropy of the total system 12, subsystem 1 and subsystem 2, respectively. Since \eqref{pseudo_additive_of_T_q} is nothing but the $q$-sum \eqref{A.6}, with the entropies measured in units of $k$, i.e.\ $T_q^{12} = T_q^1 \oplus_q T_q^2$, we shall refer to this property as a \emph{$q$-additivity} throughout this paper. The ordinary additivity is recovered for $q=1$, so that $1-q$ measures the deviation from it. Under the limit $q\rightarrow 1$ of \eqref{Tq}, one obtains}
\begin{equation}
    S_{BG} \, = \, -k\sum_i p_i \ln\, p_i \;,
\end{equation}
which is the standard Boltzmann-Gibbs entropy satisfying \( S_{BG}^{12} \, = \,  S_{BG}^1 +  S_{BG}^2\) with the product rule of the joint probability.
There is another one-parameter entropy known as the Rényi entropy
\begin{equation}
    R_q \, = \, \frac{k}{1-q} \ln \sum_i p_i ^q \, , \hspace{0.5cm} q > 0\;.
\end{equation}
One of the interesting features is the additivity
\begin{equation}
     R_{q}^{12} \, = \,  R_{q}^1 +  R_{q}^2 \;,
\end{equation}
which seems to share this feature with the Boltzmann-Gibbs entropy. Moreover, there is the relation between Tsallis entropy and  Rényi entropy as
\begin{equation}
    T_q \, = \, \frac{k}{1-q} [e^{(1-q)R_q/k} -1] \;,
\end{equation}
or
\begin{equation}
    R_q \,=\, \frac{k}{1-q}\ln[1+(1-q)T_q/k] \;.
\end{equation}
Applying \eqref{q-expo} and \eqref{q-log} in the appendix, we can write
\begin{equation}
    T_q \,=\, k\,\ln_q\, (e^{R_q/k}) \;,
\end{equation}
\begin{equation}
     R_q \,=\, k\,\ln\, (e_q^{T_q /k}) \;.
\end{equation}
Using \eqref{q-expo_of_q-log} and \eqref{q-log_of_q-expo}, one has
\begin{equation}
    e_q^{T_q/k} \, = \, e^{R_q/k}
    \label{eT=eR}\;.
\end{equation}
Now, we are ready to explore the properties of \(q\) deformed Hamiltonian. We shall first start with a bit of history. With the system of one degree of freedom and the equation of motion is given by
\begin{equation}
    m\ddot{x} = -\frac{dV}{dx}
    \label{F=ma}\;,
\end{equation}
it is not difficult to work out that the Lagrangian for this system is $L_N(\dot x,x)=T(\dot x)-V(x)$, where $T(\dot x)=m\dot x^2/2$ is the kinetic energy and $V(x)$ is the potential energy. However, from the non-uniqueness property of the Lagrangian, one can multiply and add constants such that \(\, L_N \to \beta L_N + \alpha \, \) without affecting the equation of motion \eqref{F=ma}. Moreover, the total derivative term \( 
\frac{df(x,t)}{dt}\) can be directly added into the Lagrangian and, again, this process will not alter the equation of motion. This is quite trivial since the total derivative term becomes the boundary term and does not play a part in the variational process of the action leaving the same Euler-Lagrange equation.
\\
\\
In \cite{MultiL}, a simple question was addressed: Are there other forms of Lagrangian producing the equation of motion \eqref{F=ma}? It turns out to be ``YES". A new form of the Lagrangian called the multiplicative form \(L(\dot x,x)=\, F(\Dot{x})G(x)\) is proposed, where $F(\dot x)$ and $G(x)$ are to be determined. Employing the inverse problem of the calculus of variations together with the constraint \eqref{F=ma}, the multiplicative Lagrangian is obtained
\begin{equation}\label{LL}
    L_\lambda(x, \Dot{x}) = m\lambda^2 \left( e^{-\frac{\Dot{x}^2}{2\lambda^2}} + \frac{\Dot{x}^2}{\lambda^2} \int_0^{\Dot{x}} e^{-\frac{{v}^2}{2\lambda^2}} \, dv \right) e^{-\frac{V(x)}{m\lambda^2}}-m\lambda^2\;,
\end{equation}
where $\lambda$ is a parameter in the velocity unit. It is not difficult to see that this new form of the Lagrangian gives exactly the same equation of motion \eqref{F=ma} and, under the limit $\lambda \rightarrow \infty$, the standard Lagrangian $\lim_{\lambda \rightarrow \infty}L_{\lambda}=L_N$ is recovered. Therefore, \eqref{LL} can be treated as a one-parameter version of the standard Lagrangian.
\\
\\
Applying the Legendre transformation
\begin{equation}
    H_\lambda(p,x) \, \equiv \, p_\lambda\dot{x} - L_\lambda(\dot{x},x) \;,\;\;\text{where}\;\;p_\lambda=\frac{\partial L_\lambda}{\partial \dot{x}}\;,
\end{equation}
one obtains
\begin{equation}\label{HL}
    H_\lambda (p,x) \, = \,  -m\lambda^2 e^{-\frac{H_N}{m\lambda^2}}+ m\lambda^2 \;.
\end{equation}
Here  \( \, H_N(p,x) \, =\, \frac{p^2}{2m} + V(x)\,\) is the standard Hamiltonian. Again, it is not difficult to see that this new Hamiltonian gives exactly the same equation of motion \eqref{F=ma} and also under the limit $\lim_{\lambda \rightarrow \infty}H_{\lambda}=H_N$ is recovered.
Now, if we reparameterise the parameter such that \(\, -m\lambda^2 \,=\, \frac{\gamma}{1-q}  \,\), where \(\gamma\) is a constant in the unit of energy and $q\in \mathbb R$. Then the multiplicative Hamiltonian \eqref{HL} becomes
 \begin{equation}
     H_q(p,x) = \frac{\gamma}{1-q}[e^{(1-q)H_N/\gamma}-1]
     \label{H_q} \;,
 \end{equation}
which is now called a $q$-deformed Hamiltonian. Consequently, if we have \(H_N^{12} \, = \,H_N^{1} + H_N^{2} \,\), the $H_q^{12}$ will be
 \begin{equation}\label{Hq1}
     H_q^{12} \, = \, H_q^1 + H_q^2 + (1-q)\frac{H_q^1 H_q^2}{\gamma}\; ,
 \end{equation}
 which obviously exhibits the non-additive property as those given in \eqref{pseudo_additive_of_T_q} for Tsallis entropy. In the limit $q\rightarrow 1$, \eqref{Hq1} becomes a standard additive between two Hamiltonian: $H_1^{12}=H_1^1+H_1^2=H_N^{12}$.
 \\
 \\
 Next, we can express \eqref{H_q} in terms of $q$-exponential \eqref{q-expo} and $q$-logarithm \eqref{q-log}, resulting in
 \begin{equation}
     H_q\, =\, \gamma\, \ln_q\, e^{H_N / \gamma},
     \label{eq1.21} 
 \end{equation}
 or
 \begin{equation}
     e_q^{\oplus_q H_q/\gamma} \, = \, e^{+H_N/\gamma}
     \label{eq1.22} .
 \end{equation}
\\
 With properties of the $q$-deformed Hamiltonian, it seems to suggest us that there must be a connection between Tsallis entropy and $q$-deformed Hamiltonian from those hints of either the \(q\)-additive features \eqref{pseudo_additive_of_T_q} and \eqref{Hq1} or the relations \eqref{eT=eR} and \eqref{eq1.22}. But the connection between entropy and Hamiltonian is quite not straightforward since it depends on the structure of the system. From \eqref{eq1.22}, we can interpret the left-hand side of the equation as the Tsallis power law distribution while the right-hand side seems to give the energy distribution in statistical mechanics. However, \cite{10.1063/1.5111333} shows that Boltzmann-Gibbs entropy and thermodynamic entropy, in general, they are not directly connected but only special cases where a system is in equilibrium with the reservoir known as a canonical ensemble.
 \\
 \\
 Then, we are interested in the property of a canonical ensemble using the \(q\)-deformed Hamiltonian and find its corresponding entropy. This approach is not the same with MSE \cite{PhysRev.106.620} which obtained the probability distribution or density from entropy with some constraints. Here, we rather start from the first principle to construct a statistical ensemble for probability density and use it to find the entropy. We note that this kind of strategy from the first principle has been done before in \cite{OU20085761} and \cite{DEALMEIDA2005395,ALMEIDA2001424} with the standard Hamiltonian leading to the imperfect reservoir condition. The difference is that we take the reservoir to be perfect as one normally does in standard textbooks of statistical mechanics together with the help of $q$-algebra to construct the ensemble.

\section{Statistical ensembles with \(H_q\)}
\label{Construct_ensemble}
\subsection{The definition of \(q\)-version of phase space density matrix}
\label{q_Hamiltonian_system}
To construct a statistical ensemble with the $q$-deformed Hamiltonian, we first construct a microcanonical ensemble for an isolated system. In the case of a standard Hamiltonian, the phase space density matrix of the system is given by
\begin{equation}
    \rho_{N} \,=\, \frac{\delta(E_N-H_N)}{Z_N}\;,
\end{equation}
where $\delta(x)$ is a delta function and $Z_N$ is a normalization constant. Here, a label \(N\) denotes for the standard energy in the Newtonian context.
For the $q$-deformed Hamiltonian case, it is very natural to define the phase space density matrix in the same sense such that
\begin{equation}
    \rho_{q} \,=\, \frac{\delta(E_q-H_q)}{Z_q} \;,
\end{equation}
where $Z_q$ is a normalization constant in this case. 
However, the issue arises when we consider the phase space density matrix of a large system \(H_q^{12}\). According to \(q\)-additive  feature, one has \(H_q^{12} = H_q^{2} \mathbin{\oplus_q} H_q^{2}\). Here is a point, in the case of standard Hamiltonian, with $H_N^{12}$, one can find the Hamiltonian of subsystem 1 trivially: $H_N^1 = H_N^{12} - H_N^2$. One can process the same with $H_q^{12}$ and would get $H_q^1 \neq H_q^{12} - H_q^2=H_q^{1} \mathbin{\oplus_q} H_q^{2}- H_q^2$. 
However, this problem can be fixed by introducing a factor such that
\begin{equation}
\frac{H_q^{12} - H_q^{2}}{1 + (1-q)H_q^{2}/\gamma} = H_q^{1}\;.   
\end{equation}
With \eqref{A.7}, we now arrive
\begin{equation}
    H_q^{1} = H_q^{12} \mathbin{\ominus_q} H_q^{2}
\end{equation}
and the phase space density for system $12$ becomes
\begin{equation}
    \rho_{q}^{12} \,=\, \frac{\delta(E_q^{12}-(H_q^{1} \mathbin{\oplus_q} H_q^{2}))}{Z_q}.
\end{equation}
So far, we have fixed one problem and another problem to go with the probability property of the $\rho_q^{12}$.
The main issue directly comes from tracing out any subsystem since there is an extra term \((1-q)H_q^1 H_q^2 /\gamma\), see \cite{OU20085761}.
But if we treat the parameter \(q\) as a measure of the degree of \(q\)-additive, we can generalise the standard subtraction (between \(E_q^{12}\) and \(H_q^{12}\)) into the $q$-algebra context. By doing this, it is easy to proceed with the calculation from the property of $q$-deformed algebra \eqref{A.19}, which looks very supportive and more natural to do operation than the standard one.
We conclude that, with a \(q\)-deformed Hamiltonian, one should consider \(E \mathbin{\ominus_q} H\). Then, our phase space density can be expressed in the form
\begin{equation}
    \rho_{q}^{12} \,=\, \frac{\delta_q(E_q^{12} \mathbin{\ominus_q} H_q^{12})}{Z_q}.
\end{equation} 
The last point is that one needs to use the \(q\)-Dirac delta function and a reason shall be revealed in the next section.
\subsection{Properties of \(\rho_q^{12}\)}
\label{Two_ststem_contact}
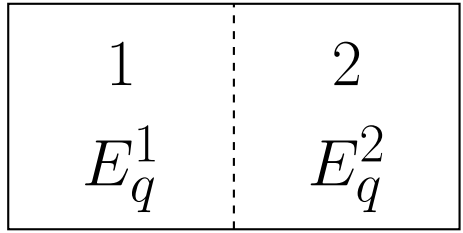
\begin{figure}[h]
    \centering
    \begin{tikzpicture}

    \fill[white] (-0.5,-0.5) rectangle (6.5,3.5); 
    
    \fill[white!20] (0,0) rectangle (6,3);

 \draw[thick, black] (0,0) rectangle (6,3);

    \draw[dashed, thick] (3,0) -- (3,3);         

    \node at (1.5, 2.2) {\Huge 1};           
    \node at (4.5, 2.2) {\Huge 2};           
    \node at (1.5, 0.8) {\Huge $E_q^1$};     
    \node at (4.5, 0.8) {\Huge $E_q^2$};     

    \end{tikzpicture}
    \caption{An isolated system composed of subsystems 1 and 2.}
    \label{fig:isolate-system}
\end{figure}

In this section, we shall explore the properties of the phase space density matrix. To achieve the goal, we examine an isolated composite system, with a fixed energy \( E_q^{12} \), which consists of subsystem 1 with energy $E_q^1$ and subsystem with energy $E_q^2$, see figure \ref{fig:isolate-system}.

The phase space density matrix of this composite system  \( \rho_q^{12} \), in a microcanonical ensemble, is given by
\begin{equation}\label{rhoq2}
    \rho_q^{12} \equiv \frac{\delta_q (E_q^{12} \mathbin{\ominus_q} (H_q^1 \mathbin{\oplus_q} H_q^2))}{\Omega_q^{12} (E_q^{12})}\;,
\end{equation}
where \( \Omega_q^{12} (E_q^{12}) \) is a normalisation factor. From the definition of probability 
\begin{equation}
    1 = \int_{\Sigma_{12}} d\Gamma_{12}  \rho_q^{12},
\end{equation}
where $\Sigma_{12}$ is the phase space and $d\Gamma_{12}=dp_1dx_1dp_2dx_2$ is an element hyper-volume, thus, the normalisation factor is
\begin{equation}
    \Omega_q^{12} (E_q^{12}) = \int_{\Sigma_{12}} d\Gamma_{12} \, \delta_q (E_q^{12} \mathbin{\ominus_q} (H_q^1 \mathbin{\oplus_q} H_q^2)).
\end{equation}
Suppose we are interested in examining subsystem 1. What we need to do is to trace out subsystem 2 resulting in a reduced phase space density matrix  
\begin{equation}
    \rho_q^1(p_1, x_1) = \int_{\Sigma_2} d\Gamma_2 \, \rho_q^{12} (p_1, x_1, p_2, x_2)=\frac{\int_{\Sigma_2} d\Gamma_2 \, \delta_q (E_q^{12} \mathbin{\ominus_q} (H_q^1 \mathbin{\oplus_q} H_q^2))}{\Omega_q^{12} (E_q^{12})}\;.
\end{equation}
Calling the property of the \(q\)-difference \eqref{A.7} \[
E_q^{12} \mathbin{\ominus_q} (H_q^1 \mathbin{\oplus_q} H_q^2) = (E_q^{12} \mathbin{\ominus_q} H_q^1) \mathbin{\ominus_q} H_q^2\;,
\]
where \( E_q^{12} \mathbin{\ominus_q} H_q^1 \) represents the energy of subsystem 2, the phase space density of subsystem 1 becomes
\begin{equation}
   \rho_q^1(p_1, x_1) \, \equiv \frac{\Omega_q^2 (E_q^{12} \mathbin{\ominus_q} H_q^1)}{\Omega_q^{12}(E_q^{12})}\;,
\end{equation}
where
\begin{equation}
    \Omega_q^2 (E_q^{12} \mathbin{\ominus_q} H_q^1) = \int_{\Sigma_2} d\Gamma_2 \, \delta_q ((E_q^{12} \mathbin{\ominus_q} H_q^1) \mathbin{\ominus_q} H_q^2)\;.
\end{equation}
Similarly, if we trace out subsystem 1, we obtain
\begin{equation}
    \rho_q^2(p_2, x_2) \equiv \frac{\Omega_q^1 (E_q^{12} \mathbin{\ominus_q} H_q^2)}{\Omega_q^{12}(E_q^{12})}\;,
\end{equation}
where
\begin{equation}
    \Omega_q^1 (E_q^{12} \mathbin{\ominus_q} H_q^2) = \int_{\Sigma_1} d\Gamma_1 \, \delta_q ((E_q^{12} \mathbin{\ominus_q} H_q^2) \mathbin{\ominus_q} H_q^1)\;.
\end{equation}
The relation between \( \Omega_q^{12} \), \( \Omega_q^1 \) and \( \Omega_q^2 \) is analogous to the convolution relation involving \(q\)-integration \eqref{q_integral} such that
\begin{equation}
    \Omega_q^{12}(E_q^{12}) \equiv \int_{(q)} d_q E_q^1 \, \Omega_q^1(E_q^1) \, \Omega_q^2(E_q^{12} \mathbin{\ominus_q} E_q^1).
\end{equation}
\textbf{Proof:} We consider
\[
\int_{(q)} d_q E_q^1 \, \Omega_q^1(E_q^1) \, \Omega_q^2(E_q^{12} \mathbin{\ominus_q} E_q^1) = \int_{(q)} \frac{d E_q^1 }{1+(1-q)E_q^1} \left\{ \int_{\Sigma_1} d\Gamma_1 \, \delta_q (E_q^1 \mathbin{\ominus_q} H_q^1) \right.\]
\[\left.\;\;\;\;\;\;\;\;\;\;\;\;\;\;\;\;\;\;\;\;\;\;\;\;\;\;\;\;\;\;\;\;\;\;\;\;\;\;\;\;\;\;\;\;\times
\int_{\Sigma_2} d\Gamma_2 \, \delta_q ((E_q^{12} \mathbin{\ominus_q} H_q^2) \mathbin{\ominus_q} E_q^1) \right\}.
\]
Reorganising the integrals and expanding the \(q\)-difference, we obtain
\begin{equation}
   \Omega_q^{12}(E_q^{12}) = \int_{\Sigma_{12}} d\Gamma_{12} \left\{ \int dE_q^1 \, \delta_q \left( \frac{E_q^1 - H_q^1}{1+(1-q)H_q^1}\right) \, \delta_q \left(\frac{(E_q^{12} \mathbin{\ominus_q} H_q^2) - E_q^1}{1+(1-q)E_q^1}\right) \left(\frac{1}{1+(1-q)E_q^1}\right)\right\}.
\end{equation}
Introducing \( u = \frac{E_q^1 - H_q^1}{1+(1-q)H_q^1} \), the integral becomes
\begin{equation}
  \Omega_q^{12}(E_q^{12}) = \int_{\Sigma_{12}} d\Gamma_{12} \left\{\int du \, \delta_q \left(\frac{(E_q^{12} \mathbin{\ominus_q} H_q^2) - E_q^1}{1+(1-q)E_q^1}\right)\left(\frac{1+(1-q)H_q^1}{1+(1-q)E_q^1}\right) \delta_q (u)\right\}.
\end{equation}
Next, we employ the property of the \(q\)-Dirac delta function\footnote{At \( u = 0 \) is nothing but \( E_q^1 = H_q^1 \).}. Thus, the integral will be further simplified as
\begin{equation}
    \Omega_q^{12}(E_q^{12})  = \int_{\Sigma_{12}} d\Gamma_{12} \, \delta_q \left(\frac{(E_q^{12} \mathbin{\ominus_q} H_q^2) - H_q^1}{1+(1-q)H_q^1}\right)\left(\frac{1+(1-q)H_q^1}{1+(1-q)H_q^1}\right).
\end{equation}
Applying the definition of \(q\)-difference, we obtain
\begin{equation}
    \int_{\Sigma_{12}} d\Gamma_{12} \, \delta_q \left(\frac{(E_q^{12} \mathbin{\ominus_q} H_q^2) - H_q^1}{1+(1-q)H_q^1}\right)\left(\frac{1+(1-q)H_q^1}{1+(1-q)H_q^1}\right) = \int_{\Sigma_{12}} d\Gamma_{12} \, \delta_q ((E_q^{12} \mathbin{\ominus_q} H_q^2) \mathbin{\ominus_q} H_q^1).
\end{equation}
Finally, this integral represents the microstate of the composite system. Therefore, this mathematical structure suggests that the phase space density matrix in this context must be in the form \eqref{rhoq2} involving $q$ parameter.

\subsection{Canonical Ensemble}
\label{Derivation_pseudo_canonial}
In this section, we shall formulate the canonical ensemble in the context of the $q$-deformed Hamiltonian, see the standard one \cite{swendsen2012,Schwabl2006}. The procedure will be the same as the standard method by treating subsystem 2 as a reservoir and focusing on subsystem 1. From the previous section, we recall
\begin{equation}
    \rho_{q}^{1}(p_1, x_1) = \frac{\Omega_{q}^{2} (E_q^{12} \mathbin{\ominus_{q}} H_q^{1})}{\Omega_{q}^{12}(E_q^{12})}\;.
\end{equation}
We then expand \( \ln_q\Omega_{q}^{2} (E_q^{12} \mathbin{\ominus_{q}} H_q^{1}) \) with the assumption \( \frac{H_q^{1}}{E_q^{12}} \ll 1 \). At this stage, one needs to use the $q$-derivative \eqref{q_derivative}, by replacing \( x = E_q^{12} \), \( y = E_q^{12} \mathbin{\ominus_{q}} H_q^{1} \), and \( F = \ln_{q} \Omega_{q}^{2} \), resulting in
\begin{equation}
    \lim_{E_q^{12} \mathbin{\ominus_{q}} H_q^{1} \to E_q^{12}} \frac{\ln_{q} \Omega_{q}^{2}(E_q^{12}) - \ln_{q} \Omega_{q}^{2}(E_q^{12} \mathbin{\ominus_{q}} H_q^{1})}{H_q^{1}} = \left[1 + (1-q)\frac{E_q^{12}}{\gamma}\right] \frac{d}{dE_q^{12}} \left( \ln_{q} \Omega_{q}^{2}(E_q^{12}) \right).
\end{equation}
In the case of a standard derivative, if one ignores the limit, the first-order Taylor series will be obtained. Analogously, if we compare the \(q\)-derivative with the standard one and ignore the limit, we obtain the first-order expansion of \( \ln_{q} \Omega_{q}^{2}(E_q^{12} \mathbin{\ominus_{q}} H_q^{1}) \) as
\begin{equation}\label{T1}
    \ln_{q} \Omega_{q}^{2}(E_q^{12} \mathbin{\ominus_{q}} H_q^{1}) \, \approx \, \ln_{q} \Omega_{q}^{2}(E_q^{12}) + \left[ 1 + (1-q) \frac{E_q^{12}}{\gamma} \right] \frac{d}{dE_q^{12}} \left( \ln_{q} \Omega_{q}^{2}(E_q^{12}) \right) (-H_q^{1})\;.
\end{equation}
We note here that the parameter \( \gamma \) will be determined and interpreted later. Moreover, it is not difficult to see that under the limit $q\rightarrow 1$, \eqref{T1} will be reduced to the standard one. 
\\
\\
Taking the \(q\)-exponential and using its properties, we obtain
\begin{equation}
    \Omega_{q}^{2}(E_q^{12} \mathbin{\ominus_{q}} H_q^{1}) = \Omega_{q}^{2}(E_q^{12}) \mathbin{\otimes_{q}} e^{- \left[ 1 + (1-q) \frac{E_q^{12}}{\gamma} \right] \frac{d}{dE_q^{12}} \left( \ln_{q} \Omega_{q}^{2}(E_q^{12}) \right) H_q^{1}}_{q}\;.
\end{equation}
Using the identity \( A \mathbin{\otimes_{q}} e^x_{q} = A e^{x A^{q-1}}_{q} \), we can define \( \beta_{q} \) as
\begin{equation}
    \beta_{q} \equiv \left[ 1 + (1-q) \frac{E_q^{12}}{\gamma} \right] \frac{d}{dE_q^{12}} \left( \ln_{q} \Omega_{q}^{2}(E_q^{12}) \right) [\Omega_{q}^{2}(E_q^{12})]^{q-1}\;.
\end{equation}
which shall be treated as a $q$-deformed of the $\beta$ in the standard context\footnote{The $\gamma$ is nothing but the $kT$.}. Thus, we now have
\begin{equation}
    \Omega_{q}^{2}(E_q^{12} \mathbin{\ominus_{q}} H_q^{1}) \,= \,\Omega_{q}^{2}(E_q^{12})  e^{- \beta_{q} H_q^{1}}_{q}\;.
\end{equation}
Consequently, the phase space density of subsystem 1 is given by
\begin{equation}
    \rho_{q}^{1}(p_1, x_1) = \frac{e^{- \beta_{q} H_q^{1}}_{q}}{Z_{q}}
    \label{rho_q^12=}\;,
\end{equation}
where
\begin{equation}
    Z_{q} \equiv \frac{\Omega_{q}^{2}(E_q^{12})}{\Omega_{q}^{12}(E_q^{12})}\;.
\end{equation}
The partition function \( Z_{q} \) is defined as a normalization factor to ensure that the integral over the phase space density is 1
\begin{equation}
    Z_{q} = \int_{\Sigma} d\Gamma \, e^{- \beta_{q} H_q}_{q}.
\end{equation}
For our convenience, we will drop the subscript `1' since it is understood that our system is subsystem 1 and subsystem 2 is nothing but the reservoir.

\section{Candidate entropy and system at thermal equilibrium}
\label{Entropy_Thermal_equilibrium}
\subsection{Candidate entropy}
\label{first_entropy}
By constructing the canonical ensemble with \(q\)-deformed Hamiltonian and obtaining the phase space density matrix for the system and subsystems, we now set a task to derive the entropy. The question arises: Is the second law of thermodynamics still applicable? The answer is yes. The ensemble average of the \(q\)-deformed Hamiltonian over phase space yields the internal energy as
\begin{equation}
    U_q = \int_{\Sigma} d\Gamma \, \rho_q H_q\;.
\end{equation}
From \cite{ALMEIDA2001424,DEALMEIDA2005395}, we take the variation of \( U \) resulting in
\begin{equation}
    \delta U_q = \int_{\Sigma} d\Gamma \, (\delta\rho_q) H_q + \int_{\Sigma} d\Gamma \, \rho_q (\delta H_q)
    \label{delta_U_q}\;.
\end{equation}
The first law of thermodynamics reads
    $\delta U = \delta Q + \delta W$. 
Then, the second term on the right-hand side in \eqref{delta_U_q} corresponds to \( \delta W \). We, therefore, can identify the first term on the right-hand side of  \eqref{delta_U_q} as 
\begin{equation}
    \delta Q_q = \int_{\Sigma} d\Gamma \, (\delta\rho_q) H_q\;.
\end{equation}
According to the second law of thermodynamics, we have
\begin{equation}
    \delta S = \frac{\delta Q}{T}\;,
\end{equation}
where $T$ is the temperature. We can then find the candidate entropy function \( \altmathcal{S} \) that satisfies the second law of thermodynamics as follows
\begin{equation}
    \altmathcal{S} = \frac{1}{T} \int_{\Sigma} \altmathcal{B} \, d\Gamma + C \;,
\end{equation}
where $C$ is constant of integration and will be fixed later. The function $\mathcal{B}$ is given by
\begin{equation}
    \mathcal{B} = \int_{0}^{\rho} H_q(\zeta) \, d\zeta\; .
\end{equation}
%
From the results of the previous section, taking the \(q\)-logarithm of both sides of \eqref{rho_q^12=}, we obtain 
\begin{equation}
    \ln_q \rho_q = -\beta_q H_q\mathbin{\ominus_q} \ln_q Z_q \;.
\end{equation}
Next, we express the \(q\)-deformed Hamiltonian as a function of the phase space density matrix
\begin{equation}
    H_q(\rho_q) = -\frac{1}{\beta_q} \left[ [1 + (1-q) \ln_q Z_q] \ln_q \rho_q + \ln_q Z_q \right]\; .
\end{equation}
Substituting \( H_q(\rho_q) \) into \( \mathcal{B} \), we get
\begin{equation}
    \mathcal{B} = -\frac{(Z_q)^{1-q}}{\beta_q (1-q)} \left[ \frac{\rho^{2-q}}{2-q} - \rho \right] - \frac{\rho \ln_q Z_q}{\beta_q} \;.
\end{equation}
To avoid the confusion of the index \(q\) between subscript and superscript, we drop it since the phase space density matrix is still depend on parameter \(q\). Therefore, our candidate entropy function becomes
\begin{equation}
    \mathcal{S} = -\frac{\ln_q Z_q}{T\beta_q} - \frac{(Z_q)^{1-q}}{T\beta_q (1-q)} \int_\Sigma \left[ \frac{\rho^{2-q}}{2-q} - \rho \right] d\Gamma + C\; .
\end{equation}
By reparametrising the index \( q' = 2 - q \) and using the fact that
\begin{equation}
    \frac{\rho^{q'}}{q'} - \rho = \frac{\rho^{q'} - \rho}{q'} + \frac{\rho (1 - q')}{q'} \;,
\end{equation}
the candidate entropy function becomes
\begin{equation}
    \mathcal{S} = \frac{(Z_{2-q'})^{q'-1}}{q' T \beta_{2-q'}} \int_\Sigma \left[ \frac{\rho - \rho^{q'}}{q' - 1} \right] d\Gamma + \frac{1 - \ln_{2-q'} Z_{2-q'}}{q' T \beta_{2-q'}} + C .
\end{equation}
The last two terms vanish if we choose \( C = - \frac{1 - \ln_{2-q'} Z_{2-q'}}{q' T \beta_{2-q'}} \). Moreover, the integral 
\[
 T_{q'}\equiv\int_\Sigma \left[ \frac{\rho - \rho^{q'}}{q' - 1} \right] d\Gamma 
\]
is nothing but the Tsallis entropy. Therefore, the candidate entropy function reduces to
\begin{equation}
    \mathcal{S}_{q'} = \frac{(Z_{2-q'})^{q'-1}}{q' T \beta_{2-q'}} T_{q'}
    \label{S_q'} .
\end{equation}
We observe that if \( \frac{(Z_{2-q'})^{q'-1}}{q' T \beta_{2-q'}} \) is constant, our candidate entropy function has the same behavior as Tsallis entropy. We next need to verify whether \( \beta_{2-q'} \) is constant in thermal equilibrium.

\subsection{Thermal equilibrium condition: A property of \(\beta_q\)}
\label{Thermal_equilibrium_condition}
To analyse the thermal equilibrium condition for the system with the \(q\)-deformed Hamiltonian, we start to simplify the expression for \(\beta_q\) as
\begin{equation}
    \beta_q^{12} = \left[ 1+(1-q)\frac{E_q^{12}}{\gamma} \right]\left[\frac{d}{dE_q^{12}} \ln_q \Omega_{q}^{2}(E_q^{12})\right][\Omega_{q}^{2}(E_q^{12})]^{q-1}\;.
\end{equation}
Using the definition of the $q$-logarithm, the $\beta_q$ can be expressed in the form
\begin{equation}
    \beta_q^{12} = \left[ 1+(1-q)\frac{E_q^{12}}{\gamma} \right] \frac{d}{dE_q^{12}} \ln \Omega_{q}^{2}(E_q^{12})\;.
\end{equation}
For the standard Hamiltonian case \cite{swendsen2012, Schwabl2006}, the thermal equilibrium condition is derived by finding the probability density that subsystem 1 has a specific energy \(E^1\). This probability density is given by
\begin{equation}
    \mathcal{P}(E^1) = \frac{\Omega^{2}(E^{12}-E^1) \Omega^{1}{(E^1)}}{\Omega^{12}(E^{12})}\; .
\end{equation}
The most probable value of \(E^1\), denoted by \(\mathcal{E}^1\), is found by setting the derivative of \(\mathcal{P}(E^1)\) with respect to \(E^1\) to zero
\begin{equation}
    \frac{d}{dE^1} \left( \frac{\Omega^{2}(E^{12}-E^1) \Omega^{1}{(E^1)}}{\Omega^{12}(E^{12})}\right) = 0 \;.
\end{equation}
This leads to
\begin{equation}
    \left(\Omega^{2}(E^{12} - E^1) \frac{d}{dE^1} \Omega^{1}(E^1) + \Omega^{1}(E^1) \frac{d}{dE^1} \Omega^{2}(E^{12} - E^1)\right) \Bigg|_{\mathcal{E}^1} = 0 .
\end{equation}
Given that \(E^1 + E^2 = E^{12}\) (a constant), we have \(dE^1 = -dE^2\). Thus, the equation becomes
\begin{equation}
   \Omega^{2}(E^2) \frac{d}{dE^1} \Omega^{1}(E^1) \Bigg|_{\mathcal{E}^1} = \Omega^{1}(E^1) \frac{d}{dE^2} \Omega^{2}(E^2) \Bigg|_{E^{12}-\mathcal{E}^1} .
\end{equation}
The above equation is
 $\ \beta^{1} = \beta^{2}$ which is the standard thermodynamics equilibrium condition, where $\beta^{i} = \frac{d}{dE^i} \ln\Omega^{i} (E^i)$. 
 \\ 
 \\ 
 In the \(q\)-deformed Hamiltonian case, the probability density that subsystem 1 has specific energy \( E_q^1 \) should have the same form as the standard Hamiltonian case
 \begin{equation}
    \mathcal{P}_q(E_q^1) = \frac{\Omega_q^{2}(E_q^{12}\mathbin{\ominus_q}E_q^1) \Omega_q^{1}{(E_q^1)}}{\Omega_q^{12}(E_q^{12})}\;.
\end{equation}
Then, we take the derivative and set it to zero. We note that one could take the \(q\)-derivative or the standard derivative resulting in exactly identical relation 
\begin{equation}\label{CC1}
\left(\Omega_q^{2}(E_q^{12} \mathbin{\ominus_q} E_q^{1}) \frac{d}{dE_q^{1}} \Omega_q^{1}(E_q^{1}) + \Omega_q^{1}(E_q^{1}) \frac{d}{dE_q^{1}} \Omega_q^{2}(E_q^{12} \mathbin{\ominus_q} E_q^{1})\right) \Bigg|_{\mathcal{E}_q^1} = 0 .
\end{equation}
Although the composite system has a fixed energy, it does not imply \( dE_q^{1} = -dE_q^{2} \), because of the non-additive property of \(q\)-deformed Hamiltonian. To see what we have in this context, one considers
\begin{equation} \label{Eq12}
E_q^{1} + E_q^{2} + (1-q) \frac{E_q^{1} E_q^{2}}{\gamma} = E_q^{12}\;.
\end{equation}
Differentiating \eqref{Eq12} with respect to \( E_q^{1} \), we obtain
\begin{equation} \label{Eq0}
    1 + \frac{dE_q^{2}}{dE_q^{1}} + \frac{(1-q)}{\gamma} \left(E_q^{1} \frac{dE_q^{2}}{dE_q^{1}} + E_q^{2} \right) = 0\;.
\end{equation}
We rearrange the terms in \eqref{Eq0} to obtain \( \frac{dE_q^{2}}{dE_q^{1}} \), resulting in
\begin{equation}\label{C1}
    \frac{dE_q^{2}}{dE_q^{1}} = -\frac{1 + \frac{(1-q)E_q^{2}}{\gamma}}{1 + \frac{(1-q)E_q^{1}}{\gamma}}\;.
\end{equation}
Substituting \eqref{C1} into the equilibrium condition \eqref{CC1}, we arrive
\begin{equation}
    \Omega_q^{2}(E_q^{2}) \frac{d}{dE_q^{1}} \Omega_q^{1}(E_q^{1}) \Bigg|_{\mathcal{E}_q^1} - \frac{1 + \frac{(1-q)E_q^{2}}{\gamma}}{1 + \frac{(1-q)E_q^{1}}{\gamma}} \Omega_q^{1}(E_q^{1}) \frac{d}{dE_q^{2}} \Omega_q^{2}(E_q^{2}) \Bigg|_{E_q^{12} \mathbin{\ominus_q} \mathcal{E}_q^1} = 0 \;,
\end{equation}
which can be simplified further to
\begin{equation}
    \left[1 + (1-q) \frac{E_q^{1}}{\gamma}\right] \frac{d}{dE_q^{1}} \ln \Omega_q^{1}(E_q^{1}) \Bigg|_{\mathcal{E}_1} = \left[1 + (1-q) \frac{E_q^{2}}{\gamma}\right] \frac{d}{dE_q^{2}} \ln \Omega_q^{2}(E_q^{2}) \Bigg|_{E_q^{12} \mathbin{\ominus_q} \mathcal{E}_1}.
\end{equation}
Recalling the definition of \( \beta_q^{12} \), thus, we establish a thermal equilibrium condition for the $q$-deformed Hamiltonian
\begin{equation}
    \beta_q^{1} = \beta_q^{2}\;.
\end{equation}
Therefore, we conclude that the candidate entropy function \eqref{S_q'} is equivalent to the Tsallis entropy, up to a scaling factor.

\section{Connection with thermodynamic quantities }
\label{Connection_w/_non_extensive_thermo}
\subsection{The effective phase space density}
Thermodynamic phenomena in the macroscopic world are well understood as emergent behaviors arising from the underlying randomness at the microscopic level, as described by statistical mechanics. In the previous sections, we successfully establish the statistical mechanics with the $q$-deformed Hamiltonian. Now, we are going to make a connection with the thermodynamics. With the canonical ensemble defined in section \ref{Derivation_pseudo_canonial}, we shall calculate the internal energy and Helmholtz free energy. The standard definition of these quantities related with the partition function is given by
%
%
\begin{equation}\label{U11}
    U \,=\, -\frac{\partial\, \ln\, Z}{\partial \beta} \;,
\end{equation}
and
\begin{equation}
    F \,=\, -kT\,\ln\,Z\;.
\end{equation}
In the case of the $q$-deformed Hamiltonian, these quantities can also be derived using the partition function. However, an inconsistent definition of internal energy arises during the derivation process. To see this, we shall begin by examining a $q$-version of the partition function.
%
Differentiating \(Z_q\) with respect to \(\beta_q\), we obtain
\begin{equation}
    \frac{\partial}{\partial \beta_q} e^{-\beta_q H_q}_q = \frac{\partial}{\partial \beta_q} [1-(1-q)\beta_q H_q]^{\frac{1}{1-q}} = [1-(1-q)\beta_q H_q]^{\frac{q}{1-q}} (-H_q)\;.
    \label{4.4}
\end{equation}
Employing the \(q\)-deformed exponential function, one gets
\begin{equation}
    -\frac{\partial}{\partial \beta_q} e^{-\beta_q H_q}_q = H_q (e^{-\beta_q H_q}_q)^q\;.
\end{equation}
Using the above result, we have
\begin{equation}\label{ZZ}
    -\frac{\partial}{\partial \beta_q} Z_q = \int_{\Sigma} d\Gamma \, H_q (e^{-\beta_q H_q}_q)^q\;.
\end{equation}
Here, the first problem arises when multiplying by \(\frac{1}{Z_q}\) throughout \eqref{ZZ} leading to
\begin{equation}
    -\frac{\partial}{\partial \beta_q} \ln Z_q = \frac{\int_{\Sigma} d\Gamma \, H_q (e^{-\beta_q H_q}_q)^q}{Z_q}\neq\int_{\Sigma} d\Gamma \, H_q \rho_q \;.
\end{equation}
The left-hand side seems to give what we want in \eqref{U11}. But, the right-hand side of the equation is not the ensemble average of the $H_q$.
%
\\
\\
To fix this problem, we start to consider the operator $D_{(q)}$, defined in \eqref{q_derivative}, whose eigenfunction is \(e_q^x\), where \(x = -\beta_q H_q \),
\begin{equation}
    D_{(q)}e^{x}_q = e_q^x\;.
\end{equation}
To obtain \(H_q e_q^x\) on the right-hand side, we multiply by \(\frac{\partial x}{\partial \beta_q}\) to the left-hand side. Then, we have
\begin{equation}
    \frac{\partial x}{\partial \beta_q} D_{(q)}e^{x}_q = H_q e_q^x.
\end{equation}
Although this approach seems to provide a method for differentiating the partition function to obtain the internal energy. However, the second problem inevitably pops up since the operator \(\frac{\partial x}{\partial \beta_q} D_{(q)}\) is not easy to handle with the phase space integral in the partition function. 
%
%
To tackle this second problem, initially, we hypothesize that the $q$-deformed Hamiltonian should incorporate with the $q$-function and $q$-operator\footnote{This is similar to how the internal energy is obtained by differentiating the standard logarithm of the partition function in the standard case.}. We then consider\footnote{We note that a similar way to construct the internal energy has been studied in \cite{EMF_Curado_1991,EMF_Curado_1992}, but it was not widely used.}
%
%
%
\begin{equation}
     -\frac{\partial}{\partial \beta_q} \ln_q Z_q = \int_{\Sigma} d\Gamma \, H_q \left(\frac{e^{-\beta_q H_q}_q}{Z_q}\right)^q= \int_\Sigma d \Gamma \, H_q\,\rho_q^q \neq  \int_\Sigma d \Gamma \, H_q\,\rho_q\;. 
     \label{motivate1}\;
\end{equation}
This is still unfortunate, since we could not have the ensemble average of the $q$-Hamiltonian on the right-hand side. Moreover, the probability condition does not hold
\begin{equation}
    \int_\Sigma d \Gamma \,\left(\frac{e^{-\beta_q H_q}_q}{Z_q}\right)^q \neq 1\;.
    \label{neq}
\end{equation}
To settle this problem, we propose an alternative form of the partition function given by
\begin{equation}
    \tilde{Z}_q^q \,=\,  \int_\Sigma d \Gamma \,\left(e^{-\beta_q H_q}_q\right)^q\;.
\end{equation}
Then, it is not difficult to see that
\begin{equation}
     1\,=\,  \int_\Sigma d \Gamma \,\left(\frac{e^{-\beta_q H_q}_q}{\tilde{Z}_q}\right)^q\;.
\end{equation}    
Consequently, a new phase space density matrix \( \sigma_q \) is given by
\begin{equation}
     \sigma_q \,=\, \left(\frac{e^{-\beta_q H_q}_q}{\tilde{Z}_q}\right)^q
     \label{sigma_q}.
\end{equation}
The above procedure was proposed in \cite{WANG20011431} for the discrete version with the idea of incomplete normalization and the effective probability condition
\begin{equation}
    \sum_{i=1} p_i^q \,=\, 1\;, \hspace{0.5cm} q \in [0,\infty] \;.
\end{equation}
The expectation value of a quantity \(\mathbin{O}\) is given by
\begin{equation}
    \langle \mathbin{O} \rangle = \sum_{i=1} p_i^q \mathbin{O}_i .
\end{equation}
By using MSE to maximize Tsallis entropy, the probability must be in the form
\begin{equation}
    p_i \,=\, \frac{e_q^{-\beta E_i}}{\left[ \sum_i (e_q^{-\beta E_i})^q \right]^{\frac{1}{q}}}\;.
\end{equation}
Comparing \(p_i\) and \(\sigma_q\), we then define $\sigma_q \,\equiv \, \tilde{\rho}_q^q$, where
\begin{equation}
    \tilde{\rho_q}=\frac{e^{-\beta_q H_q}_q}{\tilde Z_q}\;.
\end{equation} 
We shall call \( \tilde{\rho}_q^q\) as the effective phase space density and 
\begin{equation}
    \int_{\Sigma} d\Gamma \, \Tilde{\rho}_q^q = 1 \;.
\end{equation}
Then, the $q$-version of internal energy is given by
\begin{equation} 
     U_q \,=\, \int_{\Sigma} d\Gamma \,H_q\, \tilde{\rho}_q^q \;
     \label{U_q_effective}.
\end{equation}
We see that, with the new probability density $\tilde \rho_q$ and partition function $\tilde{Z}_q$, the internal energy is consistently defined. However, before proceeding further in the calculation, we need to establish the relationship between \(Z_q\) and \(\tilde{Z}_q\). We consider in differentiating \(x\) instead of \(\beta_q\) of \eqref{4.4} resulting in
\begin{equation}
    \frac{d}{dx} Z_q = \frac{1-q}{1-q} \int_{\Sigma} d\Gamma [1+(1-q)x]^{\frac{q}{1-q}}=\tilde{Z}_q^q\;.
\end{equation}
%
%
%
Thus, the effective phase space density matrix can be expressed as
\begin{equation}
    \sigma_q = \frac{(e_q^x)^q}{(\tilde{Z}_q)^q} = \frac{\frac{d}{dx} e_q^x}{\frac{d}{dx} Z_q}.
\end{equation}
This suggests that the effective phase space density matrix can be derived by taking the derivative with respect to \(x\) of the partition function $Z_q$. We can interpret this as applying an operator to the phase space density matrix
\begin{equation}
    \frac{\frac{d}{dx} e_q^x}{\frac{d}{dx} Z_q} = \hat{\mathbin{L}} \rho_q.
\end{equation}
To determine the operator \(\hat{\mathbin{L}}\), we consider the derivative of \(\rho_q\) with respect to \(x\)
\begin{equation}
    \frac{d}{dx} \rho_q = \frac{d}{dx} \left( \frac{e_q^x}{Z_q} \right) = \frac{Z_q \frac{d}{dx} e_q^x - e_q^x \frac{d}{dx} Z_q}{Z_q^2}\;.
\end{equation}
This can be simplified to
\begin{equation}
    \frac{d \rho_q}{dx} = \frac{(\tilde{Z}_q)^q}{Z_q} (\sigma_q - \rho_q)\;,
\end{equation}
which can be rewritten as
\begin{equation}
    \left(\frac{1}{\frac{d}{dx} \ln Z_q}\right) \frac{d \rho_q}{dx} + \rho_q = (\tilde{\rho}_q)^q \,=\, \sigma_q.
\end{equation}
This relation represents a \textbf{Legendre Transformation}. Then, the operator \(\hat{\mathbin{L}}\) is given by
\begin{equation}
    \hat{\mathbin{L}} = \left(\frac{1}{\frac{d}{dx} \ln Z_q}\right) \frac{d}{dx} + 1\;.
\end{equation}
Therefore, the effective phase space density matrix \(\sigma_q\) can be obtained by applying the operator \(\hat{\mathbin{L}}\) to the phase space density matrix \(\rho_q\).
\\
\\
Alternatively, there is a way to consistently define the internal energy as
\begin{equation}
    U_q \,=\, \frac{\int_\Sigma d \Gamma \, H_q\,(\rho_q)^q}{\int_\Sigma d \Gamma \,(\rho_q)^q} \,=\, \frac{\int_\Sigma d \Gamma \,H_q \,\left(\frac{e^{-\beta_q H_q}_q}{Z_q}\right)^q}{\int_\Sigma d \Gamma  \,\left(\frac{e^{-\beta_q H_q}_q}{Z_q}\right)^q}\;,
    \label{U_q_esscort}
\end{equation}
which was introduced in \cite{TSALLIS1998534} with a new definition of the probability called the escort probability given by
\begin{equation}
    \sigma_q ' \,=\, \frac{(\rho_q)^q}{\int_\Sigma d \Gamma \,(\rho_q)^q}\; .
\end{equation}
Interestingly, if one carefully looks at the \eqref{U_q_esscort} and \eqref{U_q_effective}, it turns out that they both are written in terms of the same phase space density matrix. In other words, the effective phase space density and escort phase space density are equivalent in defining internal energy: \(\sigma_q \,=\, \sigma_q '\). This constraint for defining internal energy has been widely applied in various fields using the MSE method.

%
%

\subsection{ The \(q\)-additive thermodynamic quantities}
From the previous section, we settle on a way to define the $q$-version of the internal energy. Now, we are ready to further proceed in studying the properties of the thermodynamic functions, e.g. internal energy and Helmholtz's free energy. We first recall the $q$-version of the internal energy in \eqref{U_q_effective}
\begin{equation}
    U_q = \frac{\int_{\Sigma} d\Gamma \, H_q \, (e_q^{-\beta_q H_q})^q}{\int_{\Sigma} d\Gamma \, (e_q^{-\beta_q H_q})^q}\;,\nonumber 
\end{equation}
which can be expressed in terms of the $q$-version of the partition function, see also \cite{WANG20011431}, as
\begin{equation}
    U_q = -\frac{1}{\tilde{Z}_q^q} \frac{\partial}{\partial \beta_q} Z_q= -\left(\frac{1}{\frac{d}{dx} \ln Z_q} \right) \frac{\partial}{\partial \beta_q} \ln Z_q\;.
\end{equation}
%
%
%
Applying mathematical trick, the standard logarithm can be replaced by the \(q\)-logarithm resulting in
\begin{equation}
    U_q = -\left(\frac{1}{\frac{d}{dx} \ln_q Z_q} \right) \frac{\partial}{\partial \beta_q} \ln_q Z_q\;.
\end{equation}
The internal energy definitely possesses the \(q\)-additive property as a consequence of the $q$-deformed Hamiltonian, see again \cite{WANG20011431}. This \(q\)-additive property can be directly checked by replacing \( H_q^{12}\) with \(H_q^1 \mathbin{\oplus_q} H_q^2\) into the effective phase space density matrix. Then, using the independent condition \( (\tilde{\rho}_q^{12})^q = (\tilde{\rho}_q^1)^q (\tilde{\rho}_q^2)^q \), we obtain
\begin{equation}
    U_q^{12} =  U_q^{1} +  U_q^{2} + (1-q) \frac{U_q^{1}  U_q^{2}}{\gamma}   \;.
\end{equation}
The $q$-version of the Helmholtz's free energy \(F_q\) is given by
\begin{equation}
    F_q = U_q - TS_q\;,
\end{equation}
or it can also be computed from the $q$-version of the partition function 
\begin{equation}
    F_q = -kT \frac{(\tilde{Z}_q)^{q-1} - 1}{q-1}= kT \ln_q \left(\frac{1}{\tilde{Z}_q}\right)\;.
\end{equation}
%
%
%
Now, it is not difficult to work out that the $q$-version of Helmholtz's free energy exhibits non-additive property
\begin{equation}
     F_q^{12} =  F_q^{1} +  F_q^{2} + (1-q) \frac{F_q^{1}  F_q^{2}}{\gamma}\;.
\end{equation}
Other thermodynamic potentials, such as enthalpy and Gibbs free energy,  can also be determined using this framework and also have \(q\)-additive properties. These quantities can be related to each other through Legendre transformations. Therefore, these thermodynamic quantities have a \(q\)-additive property which differs from the non-extensive thermodynamics given in \cite{TSALLIS1998534} \footnote{In \cite{wang:hal-00011127}, the non-additive feature of the thermodynamic potential can be also established through the Rényi entropy.}.

\section{The candidate entropy function revisited}
\label{second_entropy}
In the previous section, the new definition of the phase space density matrix called the effective one, is derived. Then, a question naturally arises whether the candidate entropy in section \ref{first_entropy} is still valid. To answer this question, we first recalled  the internal energy \eqref{U_q_effective} from the ensemble average with the effective phase space density matrix
%
%
%
%
\[
    U_q = \int_{\Sigma} d\Gamma \, \tilde{\rho}_q^q H_q \;.
\]
We shall reprocess as what we did before again. Then, we obtain the $q$-version of the heat variation as
\begin{equation}
    \delta Q_q = \int_{\Sigma} d\Gamma \, \delta (\tilde{\rho}_q)^q H_q\;.
\end{equation}
The candidate entropy function is then given by
\begin{equation}
    \mathcal{S}_q = \frac{1}{T} \int_{\Sigma} \mathcal{\tilde{B}} \, d\Gamma + C\;,
\end{equation}
where
\begin{equation}
    \mathcal{\tilde{B}} = \int_0^{\tilde{\rho}_q} H_q(\zeta) \, q \zeta^{q-1} \, d\zeta \;.
\end{equation}
Using \eqref{sigma_q} to express the \(q\)-deformed Hamiltonian in terms of the effective phase space density matrix, we have
\begin{equation}\label{H12}
    H_q(\tilde{\rho}) = \frac{-1}{\beta_q} \left(\tilde{Z}_q^{1-q} \ln_q \tilde{\rho}_q + \ln_q \tilde{Z}_q \right) .
\end{equation}
Substituting \eqref{H12} into \(\mathcal{\tilde{B}}\), we obtain
\begin{eqnarray}\label{B11}
    \mathcal{\tilde{B}} &=& -\frac{q \tilde{Z}_q^{1-q}}{\beta_q} \int_0^{\tilde{\rho}_q} \left(\frac{\zeta^{1-q} - 1}{1-q}\right) \zeta^{q-1} \, d\zeta - \frac{q \ln_q \tilde{Z}_q}{\beta_q} \int_0^{\tilde{\rho}_q} \zeta^{q-1} \, d\zeta \;,\nonumber\\
    &=& \frac{q \tilde{Z}_q^{1-q}}{\beta_q (1-q)} \left(\frac{\tilde{\rho}_q^{q}}{q} - \tilde{\rho}_q\right) - \frac{\ln_q \tilde{Z}_q}{\beta_q} \tilde{\rho}_q^q\;.
\end{eqnarray}
%
%
%
Inserting \eqref{B11} into the candidate entropy function, we obtain
\begin{equation}
    \mathcal{S}_q = \frac{q \tilde{Z}_q^{1-q}}{\beta_q T (1-q)} \int_{\Sigma} \left(\frac{\tilde{\rho}_q^{q}}{q} - \tilde{\rho}_q\right) d\Gamma - \frac{\ln_q \tilde{Z}_q}{\beta_q T} \int_{\Sigma} \tilde{\rho}_q^q \, d\Gamma + C\; .
\end{equation}
Using the fact that
\begin{equation}
    \frac{q}{1-q} \left(\frac{\tilde{\rho}_q^{q}}{q} - \tilde{\rho}_q\right) = \tilde{\rho}_q \left(\frac{\tilde{\rho}_q^{q-1} - 1}{1-q}\right) + \tilde{\rho}_q \;,
\end{equation}
the candidate entropy function will be expressed in the form
\begin{equation}
    \mathcal{S}_q = \frac{\tilde{Z}_q^{1-q}}{\beta_q T} \int_{\Sigma} \tilde{\rho}_q \ln_q \left( \frac{1}{\tilde{\rho}_q} \right) d\Gamma + \frac{\tilde{Z}_q^{1-q}}{\beta_q T} \int_{\Sigma} \tilde{\rho}_q \, d\Gamma - \frac{\ln_q \tilde{Z}_q}{\beta_q T} \int_{\Sigma} \tilde{\rho}_q^q \, d\Gamma + C\; .
\end{equation}
Employing the calculation in \cite{WANG20011431}, we have
\begin{equation}
    \beta_q = \frac{\tilde{Z}_q^{1-q}}{kT}\; .
\end{equation}
Thus, the first term in the candidate entropy function is again nothing, but the Tsallis entropy. The integral in the second term will be defined as \(\mathcal{Q}=\int_\Sigma \tilde{\rho}_q \, d\Gamma \), see also \cite{WANG20011431}. The integration in the third term satisfies the probability condition. Then, the candidate entropy function becomes
%
%
\begin{eqnarray}
    \mathcal{S}_q &=& T_q + \frac{\tilde{Z}_q^{1-q}}{\beta_q T} \mathcal{Q} - \frac{1}{\beta_q T} \left( \frac{\tilde{Z}_q^{1-q} - 1}{1-q} \right) + C \;,\nonumber\\
    &=& T_q - \tilde{Z}_q^{q-1} \ln_q(\tilde{Z}_q e_q^{-\mathcal{Q}}) + C\;.
\end{eqnarray}
%
%
%
Here, we can choose the constant $C$ to eliminate the second term as we did in the previous derivation. Therefore, the candidate entropy function is perfectly the Tsallis entropy $ \mathcal{S}_q = T_q$.

\section{Concluding summary}
\label{summary}
In this work, we establish a new connection between Tsallis entropy and the \( q \)-deformed Hamiltonian (or non-additive Hamiltonian) through the statistical mechanics approach employing  \( q \)-algebra techniques. We emphasise here that the approach presented in this work is different from other works in the literature as we do have the $q$-deformed Hamiltonian, \textcolor{black}{which is equipped with the intriguing \(q\)- additivity as its intrinsic property}, at our disposal. Specifically, we do not begin with the assumption, that there first exists the Tsallis entropy to construct the statistical ensemble, but we start from the first principle (from scratch) with the $q$-deformed Hamiltonian resulting in a key quantity called the effective phase space density matrix. We further explore the connection between the Tsallis entropy and non-additive thermodynamics. The internal energy and Helmholtz's free energy are computed. A highlight feature of these quantities is the  \(q\)- additivity as a direct consequence of the $q$-deformed Hamiltonian embedded everywhere. Another key point is that the parameter $q$ may be used to measure the deviation of thermodynamic systems from ordinary additivity. At $q=1$, the system is strictly additive, whereas at $q=0$, it exhibits the maximal deviation. Here, we would like to provide an argument to support our idea in reinterpreting the parameter $q$. In \cite{Tsallis2023ch3}, Tsallis himself mentions that the velocity addition in special relativity doesn't have additive property
\textcolor{black}{
\begin{equation}
    v_{13}/c \,=\, \frac{(v_{12}/c)+(v_{23}/c)}{1+(v_{12}/c)(v_{23}/c)}\;.
\end{equation}
}
The reason why Einstein was willing to forgo the simple additive feature in Galilean relativity is because this is a price to pay
to keep the invariant of \(c\) for every inertia frame. Moreover, Tsallis points out that the relativistic energy-momentum equation \( E^2 \,=\, m^2c^4 + p^2c^2 \) can be rewritten in terms of $q$-exponential as
\begin{equation}\label{EE1}
    E/mc^2 \,=e_{q}^{(1/2)(p/mc)^2}= e_{q}^{(p^2/2m)/mc^2}\;,\;\;\;\text{where}\;\;q=-1\;.
\end{equation}
We now recall the $q$-deformed Hamiltonian \eqref{H_q} and define \(\frac{\gamma}{1-q}\,=\,H_0\). Then, the $q$-deformed Hamiltonian becomes
\begin{equation}
    H_q/H_0 \,=\, e^{H_N/H_0} - 1\;,
\end{equation}
The term $-1$ can be removed without affecting physics resulting in 
\begin{equation}\label{H112}
    H_q/H_0 \,=\, e^{H_N/H_0} \;.
\end{equation}
We point out again here that $H_q^{12}/H_0\neq H_q^1/H_0+H_q^2/H_0$, reflecting a non-additive feature. We propose a $q$-deformed version of \eqref{H112} as
\begin{equation}\label{H113}
    \mathcal{H}_q/H_0 \,=\, e_q^{H_N/H_0} \;,
\end{equation}
which is the same with \eqref{EE1} when \(q\) equal to \(-1\). Given $H_N^{12}=H_N^1+H_N^2$, we have
\begin{equation}
\mathcal{H}_q^{12}/H_0=e_q^{(H_N^1+H_N^2)/H_0}=e_q^{H^1_N/H_0}\mathbin{\otimes_q}e_q^{H^2_N/H_0}\neq \mathcal{H}_q^{1}/H_0+\mathcal{H}_q^{2}/H_0\;.
\end{equation}
These results support treating $q$ as a measure of non-additivity, constrasting with the conventional view of $q$ as an inter-system correlation index. This distinction arises because our $q$-deformed Hamiltonian departs from the standard Newtonian form (the sum of kinetic and potential energy), see \cite{Christodoulidi_2014,Martelloni_2016,Bagchi_2016,Christodoulidi_2016,Gupta_2017,Debarshee_2018,Cirto_2018,Zhao_2022,Zhao_2023,Rodríguez_2023,Rodr_2024}.
Finally, we would like to address that the multiplicative Hamiltonian and multiplicative Lagrangian, given in \cite{MultiL}, could possibly give us more insight in physics as we could not be able to have from the standard Hamiltonian and Lagrangian, see \cite{BUKAEW202357, Supanyo:2023jkh, Supanyo2022, Supanyo2024}.

\appendix
\section*{Appendix}
\addcontentsline{toc}{section}{Appendix}
\begin{appendices}
    \section{Some properties of $q$-deformed algebra and functions}
In this appendix, we shall provide several deformed algebraic structures and functions, see \cite{article,BORGES200495}. We first give a definition of the \(q\)-exponential and \(q\)-logarithm functions which play a central role in this work
\begin{equation}
    e_q^x \equiv e_q(x) \equiv [1 + (1-q)x]^{\frac{1}{1-q}} \quad (x,q \in \mathbb{R})\;,
    \label{q-expo}
\end{equation}
\begin{equation}
    \ln_q(x) \equiv \frac{x^{1-q} - 1}{1-q} \quad (x > 0)\;.
    \label{q-log}
\end{equation}
They reduce to the standard exponential and logarithm functions under the limit \(q \to 1\)
\[
    \lim_{q \to 1} \ln_q(x) = \ln(x), \quad \lim_{q \to 1} e_q^x = e^x.
\]
One also has
\begin{equation}
    \ln_q(e_q^x) \, = \, x \;,
    \label{q-log_of_q-expo}
\end{equation}
and
\begin{equation}
    e_q^{\ln_q x} \, = \, x \;.
    \label{q-expo_of_q-log}
\end{equation}
Another important $q$-deform function is the Dirac delta function, see \cite{10.1063/1.3431981},
\begin{equation}
    \delta_q(x) = \frac{2-q}{2\pi} \int_{-\infty}^{\infty} e_q^{ikx} \, dk, \quad 1 \leq q < 2\;.
\end{equation}
Next, we give several important \(q\)-deformed operations, which exhibit the similarity to standard operations. These include
\\
\textbf{\(q\)-sum:}
\begin{equation}
   x \mathbin{\oplus_q}y = x + y + (1-q)xy\;,
   \label{A.6}
\end{equation}
\textbf{\(q\)-difference:}
\begin{equation}
    x \mathbin{\ominus_q} y = \frac{x - y}{1 + (1-q)y}, \quad 1 + (1-q)y \neq 0\;,
    \label{A.7}
\end{equation}
\textbf{\(q\)-product:}
\begin{equation}
    x \mathbin{\otimes_q} y = [x^{1-q} + y^{1-q} - 1]^{\frac{1}{1-q}}, \quad (x, y > 0)\;,
\end{equation}
\textbf{\(q\)-ratio:}
\begin{equation}
    x \mathbin{\oslash_q} y = [x^{1-q} - y^{1-q} + 1]^{\frac{1}{1-q}}\;, \quad (x, y > 0)\;.
\end{equation}
%
%
%
\vspace{0.5cm}
\textbf{Properties of \(q\)-sum}
%
%
\\
1. Commutativity:
\begin{equation}
   x \mathbin{\oplus_q} y = y \mathbin{\oplus_q} x\;,
\end{equation}
2. Associativity:
\begin{equation}
   (x \mathbin{\oplus_q} y) \mathbin{\oplus_q} z = x \mathbin{\oplus_q} (y \mathbin{\oplus_q} z)\;,
\end{equation}
3. Identity element:
\begin{equation}
   x \mathbin{\oplus_q} 0 = x\;,
\end{equation}
4. Non-distributive property:
\begin{equation}
   a(x \mathbin{\oplus_q} y) \neq ax \mathbin{\oplus_q} ay\;.
\end{equation}
\textbf{Properties of \(q\)-difference}
\\
The \(q\)-difference is defined as the inverse operation of the \(q\)-sum and introduce \(\mathbin{\ominus_q} x\) as the inverse of \(x\) under \(q\)-sum satisfying
\begin{equation}
   x \mathbin{\oplus_q} (\mathbin{\ominus_q} x) = 0\;.
\end{equation}
Therefore, the \(q\)-difference between two elements \(x\) and \(y\) can be written as
\begin{equation}
   x \mathbin{\ominus_q} y = x \mathbin{\oplus_q} (\mathbin{\ominus_q} y).
\end{equation}
Some properties of the \(q\)-difference include
\\
\\
1. Anti-commutativity:
\begin{equation}
   x \mathbin{\ominus_q} y = \mathbin{\ominus_q} y \mathbin{\oplus_q} x,
\end{equation}
\\
2. Non-distributive property:
\begin{equation}
   a(x \mathbin{\ominus_q} y) \neq ax \mathbin{\ominus_q} ay.
\end{equation}
\\
Both previous operations satisfy:
\begin{equation}
   x \mathbin{\ominus_q} (y \mathbin{\ominus_q} z) = (x \mathbin{\ominus_q} y) \mathbin{\oplus_q} z,
\end{equation}
\begin{equation}
   x \mathbin{\ominus_q} (y \mathbin{\oplus_q} z) = (x \mathbin{\ominus_q} y) \mathbin{\ominus_q} z = (x \mathbin{\ominus_q} z) \mathbin{\ominus_q} y .
   \label{A.19}
\end{equation}
\\
\textbf{Properties of \(q\)-product}
\vspace{0.25cm}
\\
\\
1. Commutative:
\begin{equation}
    x\mathbin{\otimes_q}y \, = \, y\mathbin{\otimes_q}x .
\end{equation}
\\
2. Associativity:
\begin{equation}
    x\mathbin{\otimes_q}(y\mathbin{\otimes_q}z) \, = \, (x\mathbin{\otimes_q}y)\mathbin{\otimes_q}z .
\end{equation}
\\
3. Identity element:
\begin{equation}
    x\mathbin{\otimes_q}1 \, = \, x.
\end{equation}
\\
\textbf{Properties of \(q\)-ratio}
\\
It is defined as a inverse multiplicative which is \(1\mathbin{\oslash_q}x\) so that
\begin{equation}
    x\mathbin{\otimes_q}(1\mathbin{\oslash_q}x) \, \equiv \, 1.
\end{equation}
\\
It satisfies many properties \cite{Borges2021}
\begin{equation}
     1\mathbin{\oslash_q} (1\mathbin{\oslash_q}x) =  \textcolor{black}{x} \, , \hspace{0.5cm} 0 \leq x^{1-q} \leq 2,
\end{equation}
\begin{equation}
     x \mathbin{\oslash_q} (y \mathbin{\oslash_q} z) = (x \mathbin{\oslash_q} y) \mathbin{\otimes_q} z = (x \mathbin{\otimes_q} z) \mathbin{\oslash_q} y \, , \hspace{0.5cm} z^{1-q}-1 \leq y^{1-q} \leq x^{1-q} +1 .
\end{equation}
%
%
Moreover, \(q\)-algebra can be expressed to \(q\)-logarithm and $q$-exponential
\begin{equation}
    \ln_q(xy) = \ln_q x \mathbin{\oplus_q} \ln_q y, \quad e_q(x) e_q(y) = e_q(x \mathbin{\oplus_q} y),
\end{equation}
\begin{equation}
    \ln_q(x \mathbin{\otimes_q} y) = \ln_q x + \ln_q y, \quad e_q(x) \mathbin{\otimes_q} e_q(y) = e_q(x + y),
\end{equation}
\begin{equation}
    \ln_q\left(\frac{x}{y}\right) = \ln_q x \mathbin{\ominus_q} \ln_q y, \quad \frac{e_q(x)}{e_q(y)} = e_q(x \mathbin{\ominus_q} y),
\end{equation}
\begin{equation}
    \ln_q(x \mathbin{\oslash_q} y) = \ln_q x - \ln_q y, \quad e_q(x) \mathbin{\oslash_q} e_q(y) = e_q(x - y).
\end{equation}
\vspace{0.2cm}
These relations hold under the following conditions
\begin{itemize}
   
    \item (A.24) \hspace{0.2cm} \( x > 0,\, y > 0 \),\hspace{0.2cm} and \hspace{0.2cm} \( x \geq_q 0 \) \hspace{0.1cm} or \hspace{0.1cm} \( y \geq_q 0 \),
    
    \item (A.25) \hspace{0.2cm} \( x^{1-q} + y^{1-q} \geq 1 \), \hspace{0.2cm} with \hspace{0.2cm} \( x \geq_q 0 \) \hspace{0.1cm} and \hspace{0.1cm} \( y \geq_q 0 \),
   
    \item (A.26) \hspace{0.2cm} \( x > 0,\, y > 0 \), \hspace{0.2cm} and \hspace{0.2cm} \( y >_q 0 \),
   
    \item (A.27) \hspace{0.2cm} \( x^{1-q} + 1 \, \geq \, y^{1-q} \), \hspace{0.2cm} with \hspace{0.2cm} \( x \geq 0 \) \hspace{0.1cm} or \hspace{0.1cm}  \( y \geq 0 \),
\end{itemize}
where \( x\, \geq_q \, 0 \) means \( 1 + (1-q)x \, \geq \, 0 \).
\vspace{0.5cm}
\\
\textbf{The \(q\)-Calculus}
\vspace{0.2cm}
\\
The \(q\)-derivative is defined as
\begin{equation}
    D_{(q)}f(x) \equiv \lim_{y \to x}\, \frac{f(x)-f(y)}{x \ominus_q y} \, = \, [1+(1-q)x]\frac{df(x)}{dx}\;.
     \label{q_derivative}
\end{equation}
The \(q\)-exponential is the eigenfunction of the \(q\)-derivative operator \(D_{(q)}\). The corresponding \(q\)-integral is given by:
\begin{equation}
    \int_{(q)}\, f(x) \, d_q x  \, = \, \int \frac{f(x)}{1+(1-q)x} \, dx\;,
     \label{q_integral}
\end{equation}
where
\begin{equation}
    d_q x \, \equiv \, \lim_{y \to x}\, x \ominus_q y 
 \, = \, \frac{1}{1+(1-q)x} \, dx.
\end{equation}
We note that
\begin{equation}
    \int_{(q)}\, D_{(q)}f(x)\, d_q x \, = \, D_{(q)}\int_{(q)}\, f(x)\, d_q x \,  = \, f(x).
\end{equation}
The dual derivative operator \(D^{(q)}\), corresponding to the \(q\)-derivative operator \(D_{(q)}\), is defined as

\textcolor{black}{
\begin{equation}
     D^{(q)}f(x) \equiv \lim_{y \to x}\, \frac{f(x) \ominus_q f(y)}{x-y}  \, = \, \frac{1}{1+(1-q)f(x)} \, \frac{df(x)}{dx}.
\end{equation}
}
Its corresponding dual \(q\)-integral is, see \cite{Borges2021}, 
\begin{equation}
    \int^{(q)}\, f(x)\, dx  \, = \, \int [1+(1-q)f(x)] \, f(x) \, dx .
\end{equation}
Similarly, the following relation holds
\begin{equation}
    \int^{(q)}\, D^{(q)}f(x)\, dx  \, = \, D^{(q)}\int^{(q)}\, f(x)\, dx  \, = \, f(x)\;,
\end{equation}
which is similar to the standard one.

\end{appendices}
\section*{Declarations}

\noindent\textbf{Funding} \\
This research has received funding
support from the NSRF through the Human Resources Program Management Unit
for Human Resources \& Institutional Development, Research
and Innovation [grant number 39G670016].

\vspace{0.5em}
\noindent\textbf{Availability of data and material} \\
The authors confirm that the data supporting the findings of this study are available within the article and its supplementary materials.

\vspace{0.5em}
\noindent\textbf{Competing interests} \\
The authors have no competing interests to declare that are relevant to the content of this article.

\vspace{0.5em}
\noindent\textbf{Authors Contributions} \\
Sikarin Yoo-Kong initiated and conceptualised the study, performed and verified the calculations, interpreted the results, and drafted the paper. Paradon Krisut performed most of the calculations, interpreted the results, and drafted the paper.

\bibliographystyle{unsrt}
\bibliography{references}

\end{document}